\documentclass[sigconf,10pt]{acmart}

\usepackage{tikz}
\usepackage{amsmath}
\usepackage{pdflscape}
\usepackage{xspace}
\usepackage{float}
\usepackage{verbatim}
\usepackage{amsmath}
\usepackage{bm}
\usepackage{booktabs}
\usepackage{multicol}
\usepackage{threeparttable} 
\usepackage{colortbl} 
\usepackage{graphicx} 
\usepackage{array} 
\usepackage{filecontents}
\usepackage{tikz} 
\usepackage{tabularx} 
\usepackage{algorithm}
\usepackage[noend]{algpseudocode}
\usepackage{multirow}
\usepackage{mathtools}
\usepackage{amsthm} 
\usepackage{multirow}
\usepackage{enumitem}
\usepackage{adjustbox}
\usepackage{makecell}
\usepackage{wrapfig}
\usepackage{caption}
\usepackage{subcaption}
\usepackage[compact]{titlesec}
\usepackage{booktabs}

\setlist[enumerate]{noitemsep, topsep=0pt}
\setlist[itemize]{noitemsep, topsep=0pt}
\renewcommand{\paragraph}[1]{\noindent\textbf{#1}}
\titlespacing{\paragraph}{0pt}{*0.5}{*1}
\titlespacing\section{0pt}{*1.8}{*1.1}
\titlespacing\subsection{0pt}{*1.5}{*1.1}
\titlespacing\subsection{0pt}{*1.3}{*0.8}

\renewcommand\footnotetextcopyrightpermission[1]{} 

\renewcommand\footnotetextcopyrightpermission[1]{} 
\setcopyright{none}

\acmDOI{}

\acmISBN{}

\acmConference[arXiv]{}
\acmYear{4 June 2024}

\acmPrice{}

\begin{document}
\begin{sloppypar}
	\title[Feasibility of State Space Models...]{Feasibility of State Space
	Models for Network\\Traffic Generation}
	\author{Andrew Chu}
	\authornote{Both authors contributed equally to this work.}
\affiliation{%
	\institution{University of Chicago}
}
	\author{Xi Jiang}
	\authornotemark[1]
\affiliation{%
	\institution{University of Chicago}
}

\author{Shinan Liu}
\affiliation{%
	\institution{University of Chicago}
}
		\author{Arjun Nitin Bhagoji}
\affiliation{%
	\institution{The University of Chicago}}

\author{Francesco Bronzino}
\affiliation{%
	\institution{École Normale Supérieure de Lyon}
}
\author{Paul Schmitt}
\affiliation{%
	\institution{University of Hawaii, Manoa}
}
\author{Nick Feamster}
\affiliation{%
	\institution{University of Chicago}
}
	\begin{abstract}

Many problems in computer networking rely on parsing collections of network traces,
ranging from traffic prioritization to security analysis.
Unfortunately, the availability and utility of such collections is limited due
to privacy concerns, data staleness, and low representativeness. While methods
have been developed to synthesize data to augment existing collections,
they often fall short in replicating the quality of real-world traffic. In this
paper, we i) survey the evolution of traffic simulators and generators; and ii)
propose the use of state-space models, specifically Mamba, for packet-level,
synthetic network trace generation by modeling it as an unsupervised sequence
generation problem. Preliminary evaluation indicates that,
compared to state-of-the-art, state-space models are
able to generate synthetic network traffic with 1) higher statistical similarity to
real traffic and 2) longer generation length.
Our approach thus has the potential to
reliably generate realistic and informative synthetic network traces for
downstream tasks.
	\end{abstract}

	\maketitle
	\newcommand{\chase}[1]{\textcolor{blue}{[#1]}}
	\newcommand{\andrew}[1]{{\color{orange}{\bf AC:} #1}}
	\newcommand{\shinan}[1]{\textcolor{blue}{[#1 - Shinan]}}
	\newcommand{\prs}[1]{\textcolor{green}{[#1 - PRS]}}
  \newcommand{\focus}[1]{\textcolor{red}{\bf #1}}
	\newcommand{\eg}{{\it e.g.}}
	\newcommand{\ie}{{\it i.e.}}
	\newcommand{\etal}{{et al.}}
	\newcommand{\sysname}{{NetMamba}}
	\newlength{\oldintextsep}
	\setlength{\oldintextsep}{\intextsep}
	\setlength\intextsep{0pt}

\section{Introduction}\label{sec:intro}

The increase in complexity of networked environments and the number of network
protocols and applications motivates robust and adaptable strategies for
network management and security tasks. This requires collecting and sharing
high-quality statistics, attributes/features of this data, or further, 
the raw data itself, which can be tedious or even impossible under certain constraints
\cite{sommer2010outside, mahoney2003network,mchugh2000testing,abt2014we,de2023survey}.
Worse, networks are becoming increasingly difficult to simulate via
traditional methods (\eg, NS-3~\cite{lacage2006yet}, Harpoon
\cite{sommers2004harpoon}) as the scale and complexity of networked systems
grows~\cite{kenyon2020public,ring2019survey,labovitz2010internet}.
One approach to tackle these challenges is synthetic traffic generation, which
provides a controlled method to model, simulate, and test network behaviors
for various traffic types, without the pitfalls of real-world data collection.



Current state-of-the-art (SOTA) synthetic traffic generation methods apply one
of two primary approaches.
The first generates concise sets of meta-attributes (\ie, flow statistics)
about a network trace.
These attributes capture overarching characteristics of network
flows/sessions, and are primarily used for session-level workloads  (\eg,
heuristic-based analyses, ML for networking tasks).
In contrast, the second generates complete, raw network data at the
per-packet level in packet capture (\eg, PCAP) form.
This approach, offering a comprehensive view of network communication, is often
more versatile for downstream tasks (\eg, analyses of session content,
improving classification
models).

Unfortunately, both approaches face a common challenge in that the quality of
generated data often falls short of expectation, reflected in low
statistical similarity at the raw-byte level to real network data.
Thus, generated traces may require tedious post-generation correction
to ensure protocol fidelity before being usable as either a replacement for,
or supplement to, real network data.
Motivated to
remedy these challenges,
we explore how recent work in sequence modeling might help in
generating high-quality, raw synthetic network
traffic.
%
In this paper, we treat the task of generating synthetic network traffic as a
sequence modeling problem by applying a state-space model (SSM) built on the
Mamba architecture~\cite{gu2023mamba} to networking data. Unlike transformer
models (the predominant SOTA for sequence modeling) whose time and memory
complexity scale quadratically with input length, Mamba scales
linearly. This allows our model to learn from, and generate network traces  
of both better quality, and longer duration
than the existing SOTA.
%
%
Specifically, we present the following contributions:
\begin{enumerate}
  \item \textbf{Survey of trace generation paradigms:} We provide a
  detailed overview of the evolution of synthetic traffic generators, their
  impact, and possible takeaways for motivating future trace generation work.
  \item \textbf{Apply SSMs to networking data for traffic generation:}
  We frame the task of synthetic trace generation as an unsupervised sequence
  generation problem, and train a Mamba-based SSM model from scratch on
  computer networking data to generate synthetic network traces at the
  raw-byte (\ie, PCAP) level.
  \item \textbf{Superior trace generation quality and efficiency:} We evaluate
  traces generated by our model and find they better capture complex
  intra-packet dependencies with additionally higher fidelity, than existing
  SOTA approaches. Our model pipeline further has a lower training and
  inference resource footprint than existing, comparable approaches.
\end{enumerate}

Building on our early results, we provide a roadmap for improving the
generation quality and flexibility of our model. We also discuss possible
adaptations of our model for use in various downstream networking tasks.


\begin{figure*}[th]
	\centering
	\includegraphics[width=\textwidth]{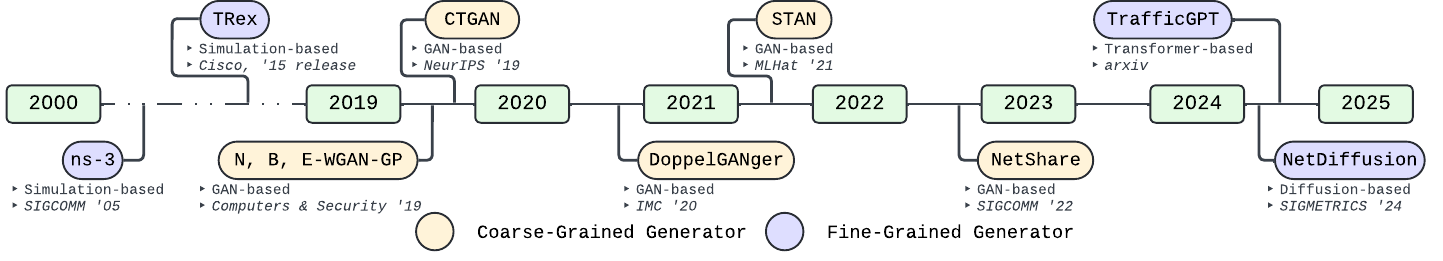}
	\caption{Timeline of synthetic network trace generation methods.}
	\label{fig:timeline}
\end{figure*}

\section{Related Work}\label{sec:related}

Early efforts in synthetic data generation such as TRex~\cite{ciscotrex2023},
NS-3~\cite{henderson2008network}, and others
\cite{buhler2022generating,vishwanath2009swing,botta2012tool,sommers2004harpoon},
are mostly simulation-based, requiring extensive human expert 
efforts to correctly build and configure. These tools can be limited as
they depend on user-defined templates and may not necessarily capture the
complex \textit{intra-flow} (\ie, relationships between
header fields across multiple packets) and \textit{intra-packet}
(\ie, header field dependencies within individual packets)
interactions reflective of complex real-world traffic. Recent methods
\cite{ring2019flow,yin2022practical,lin2020using,xu2021stan,xu2019modeling,
jiang2024netdiffusion, zhang2024trafficgpt, jiang2023generative} employ
generative ML to learn such interactions, enabling the
automatic generation of new synthetic traces. These approaches
can generally be divided into two categories: 
\textit{coarse} and \textit{fine-grained}. Figure~\ref{fig:timeline} provides a
chronological overview of the approaches discussed in this section in either
category, further discussed below.

\subsection{Coarse-Grained Generators}
Coarse-grained generators aim to synthesize meta-information or
summary statistics (\eg, flow duration and average packet sizes) about
network traffic flows (i.e., intra-flow dependencies). Such tools
\cite{ring2019flow,yin2022practical,lin2020using,xu2021stan,xu2019modeling}
mainly rely on generative adversarial networks (GANs), training two competing
neural networks against each other for synthesizing new network traces from a
training dataset. A notable example is
Lin \etal's DoppelGANger~\cite{lin2020using}, which treats modeling network 
metric traces as a sequential time-series generation task, conditioned on metadata like locations.
This
allows any single header value or intra-flow metrics to be captured for
modeling. Unfortunately, this approach scales poorly,
requiring additional training for each new value that is desired to be
captured.
Yin \etal's NetShare~\cite{yin2022practical} improves by
generating aggregate flow statistics (\eg, duration, packet
count), or packet-level header field values (\eg, time-to-live [TTL], protocol
flags). While effective, it is limited to a
small, fixed subset of high-level network attribute,
missing granular details like intra-flow dependencies (\eg, TCP options and sequence numbers)
and intra-packet dependencies (\eg, TCP header bits influencing data offset).
Thus, coarse-grained generators often
fail to capture these low-level dependencies necessary for accurately mimicking real data.

\subsection{Fine-Grained Generators}
Fine-grained generators aim to capture both intra-flow and intra-packet dependencies
of network traffic by generating complete packet captures, \ie, all raw packet bytes,
to most accurately mimic real-world traffic for downstream use.

\paragraph{Diffusion-based Approaches.}
Jiang \etal's NetDiffusion~\cite{jiang2024netdiffusion} uses image
representations of network traces with a text-to-image diffusion model to
perform fine-grained trace generation. Diffusion models are highly expressive,
resulting in synthetic traces that more accurately mimic real network
dynamics. Unfortunately, this expressiveness can produce
noisy outputs that can compromise protocol compliance.
The authors attempt to mitigate this concern using ControlNet
\cite{zhao2023unicontrolnet} to impose dependencies as
conditional controls during generation. This improves protocol
compliance, but overall is insufficient. While ControlNet may
ensure that the distribution of packet header fields values generated largely
adhere to the protocol used by the flow, it
cannot guarantee the semantic accuracy of these fields, often requiring
extensive post-generation manual corrections.

\paragraph{Transformer-based Approaches.} Transformer-based models 
\cite{vaswani2017attention} leverage the attention mechanism
\cite{bahdanau2014neural} to learn a fine-grained representation of the
semantics of their input data in an unsupervised fashion (encoder), and may
be extended to use this representation for autoregressive generation (decoder).
Many works have applied the transformer to networking tasks 
(e.g., traffic classification
\cite{dai2022shape,he2020pert,shi2023tsfn,shi2023bfcn, Lin_2022},
networking specific Q\&A~\cite{zhang2024trafficgpt}). However,
few have used the transformer for fine-grained traffic generation. Meng~\etal
\cite{meng2023netgpt} and Wang~\etal~\cite{wang2024lens} train new variants of
GPT-2 \cite{radford2019gpt2} and T5 \cite{raffel2020t5} respectively, for
coarse-grained generation at the header field value level (\ie, similar to
DoppelGANger). Qu~\etal's TrafficGPT~\cite{qu2024trafficgpt} is the most
related to our work, presenting a GPT-2 variant that generates trace sequences
up to $12{,}032$ tokens in length via various memory optimization strategies
such as linear and local window attention
\cite{katharopoulos2020transformers,xiong2021simple}, etc~\cite{peng2023rwkv,kitaev2020reformer}.
Our Mamba-based modeling approach differs from TrafficGPT
as it provides improved training time complexity
and maximum generatable context length,
which is particularly critical in downstream applications where the analysis
of flow-level attributes over extended durations is essential (\eg,
malware/intrusion detection). Additionally, traces generated by our
architecture demonstrate better quality, evidenced by a higher statistical
resemblance to real data.

\section{State Space Models and Traffic Generation}\label{sec:method}

We adapt Mamba~\cite{gu2023mamba}, a selective structured SSM, to
synthesize fine-grained network traces that capture both
intra-flow and intra-packet dependencies. We
train a new Mamba model from scratch on the raw flow contents
to generate synthetic packets of continuous traffic traces. We first
provide background information on SSMs and our motivation for using SSMs
to generate high-quality synthetic traces in Section
\ref{sub:state_space_models} and \ref{sub:whymamba}. We then provide a
technical overview of how the general, and Mamba SSM\footnote{For the rest of
the paper, we will use ``Mamba'' and ``Mamba SSM'' interchangeably.} operate in
Section~\ref{sub:mamba}.

\subsection{State-Space Models}\label{sub:state_space_models}
SSMs are probabilistic graphical models that build on the
concept of a state-space from control engineering~\cite{kalman1960new}.
Conceptually, SSMs are identical to Hidden Markov Models in objective 
(modeling discrete observations over time), but operate using
continuous, instead of discrete latent variables. Instead of attention, SSMs
encode a hidden state, representative of prior observed context of an input
sequence, using recurrent scans. Gu~\etal~\cite{gu2020hippo} and
Voelker~\etal~\cite{voelker2019legendre} show that by fixing the state
matrix used in SSMs, the encoded context can accurately and efficiently model
long-range dependencies. Gu~\etal~extend these observations in the S4
convolution kernel~\cite{gu2021s4} to make SSMs practical for training, and
more recently in Mamba~\cite{gu2023mamba}, where they introduce a
SSM with a selection mechanism, and fixed state matrix. Mamba shows strong
performance in modeling context dependent sequence generation (\ie, language
modeling), and has been extended to various domains, including
computer vision \cite{ma2024u,liu2024vmamba,xing2024segmamba}, DNA sequencing 
\cite{schiff2024caduceus}, document information retrieval
\cite{xu2024rankmamba} and speech separation \cite{li2024spmamba}.

Although general
SSMs have long existed
in the control engineering space, they have only recently been optimized
and operationalized by Gu~\etal~\cite{gu2020hippo,gu2021s4,gu2023mamba} in
the Mamba architecture to function as a possible alternative to
transformer-based approaches for sequence modeling. We find only one recent
application of Mamba to the networking domain by Wang~\etal
\cite{wang2024netmamba}, who create a variant, NetMamba, to perform traffic
classification. NetMamba claims better classification accuracy than the
existing SOTA, with improved inference time and resource
consumption. Our objective differs from NetMamba, as we
explore what modifications/adaptations are needed to use Mamba for synthesizing
raw data for network traces, rather than performing traffic inference tasks.

\subsection{Comparison with alternative methods}\label{sub:whymamba}
We chose the Mamba SSM architecture over other fine-grained generation methods
for a number of reasons.
%
Compared to diffusion models, the tokenized input used by SSMs
allows for a less processed representation of networking data. Specifically, our
model processes sequences of the decimal values for the raw bytes of flows, as
compared to an encoded image representation in NetDiffusion derived using two
steps (nPrint~\cite{nprint} intermediary format, bit-based color assignment).
Compared to transformer-based models, SSMs scale linearly (versus
quadratically) with sequence length. This allows our models to train on inputs
four times longer and generate sequences $5.5$ times longer than TrafficGPT
\cite{qu2024trafficgpt}\footnote{Training/generation hardware: NVIDIA
A40, 48GB VRAM (ours), NVIDIA Tesla V100S, 32GB VRAM (theirs).}).
This is especially useful for networking data, where even a few
seconds of communication can exceed the conventional context windows or
capacities of other methods.

\subsection{Selective Structured SSM (Mamba)}\label{sub:mamba}
Mamba, and broadly, all SSMs, use the state-space representation
introduced in control engineering by Kalman~\cite{kalman1960new}. The general
SSM uses first-order ordinary linear differential equations to capture the
relationship (output) between unobserved variables (state) and a series of
continuous observations (input), irrespective of time (\ie, is linear-time
invariant [LTI]) As the model observes more data, it encodes a representation
of the state that captures the prior context of inputs. This state is then
used to calculate an output for a given input, and can be both discretized
to be calculated
as a recurrent neural network (RNN) in linear time, and
unrolled to a convolutional neural network (CNN) for efficient training.
On this basis, Mamba implements two changes to the general SSM that provide
\textit{structure} and \textit{selection}.

For structure, the state of general SSMs suffers from numerical instability
similar to the vanishing gradient in vanilla RNNs, where successive compression
of prior state context results in poor model performance. Mamba solves this
instability by enforcing structure on the general-case SSM state matrix
(typically randomly initialized), replacing it with a HiPPO
matrix~\cite{gu2020hippo} from prior work. The HiPPO matrix introduces a
probability measure that dictates how the SSM state is compressed, improving an
SSMs ability to model long-range dependencies in sequences.

For selection, the general LTI SSM lacks expressiveness, \ie, all discrete
inputs compressed in the state, affect the state with equal weighting. In the
context of language modeling, this prevents relevant ``keywords'' from more
heavily influencing the SSM state to develop a better semantic understanding
of input. Mamba improves SSM expressiveness by removing the LTI quality of the
general SSM, and instead makes the model time-\textit{variant}, where the
state is calculated using learned (as compared to fixed) functions of the
inputs.

These structure and selection changes result in competitive
performance against transformer-based approaches for sequence
modeling with regard to generation quality, while simultaneously better
scaling (linear versus quadratic).

\section{Modeling Network Data with State-Space Models}
\label{sec:architecture}
\begin{figure}[tb]
	\centering
	\includegraphics[width=\columnwidth]{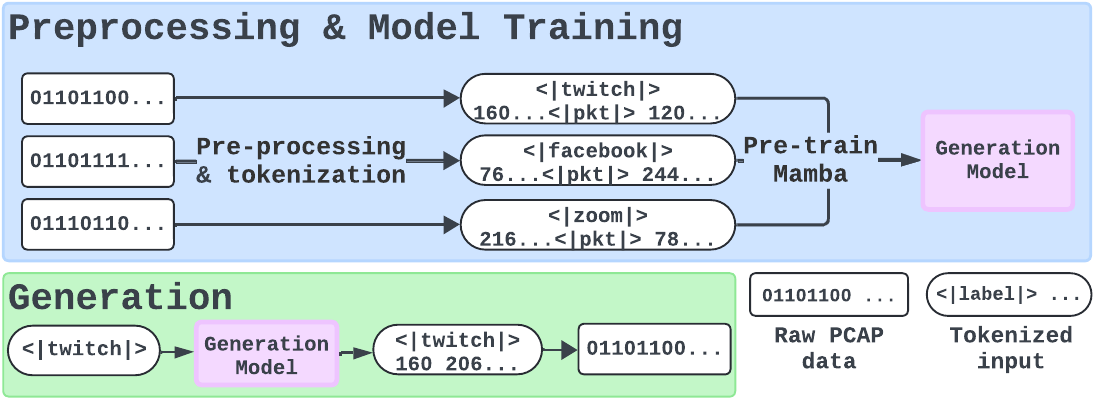}
	\caption{Model training and generation process.}
	\label{fig:pipeline}
\end{figure}

We apply the Mamba architecture to model computer networking data and
generate synthetic traces, with detailed steps below.
Our modeling pipeline (Figure~\ref{fig:pipeline}) includes
standard pre-processing and training steps for downstream use in trace
generation.

\paragraph{Trace pre-processing and tokenization.}\label{ssub:preprocessing}
We train our generation model on a tokenized representation of raw
PCAP data at the flow level. To detail, we first split each PCAP into its
comprising flows based on connection (\ie, 5-tuple of
source IP, source port, destination IP, destination port, IP protocol) using  
\texttt{pcap-splitter}~\cite{pcapsplitter}, resulting in a set of $n$ flows 
\ie, $\text{PCAP}=\{f_n\}_{n\in\mathbb{N}}$, where $f$ denotes a flow. We
then parse each packet in a flow to its sequence of raw bytes in decimal
format \ie, $\texttt{packet}=\text{Seq}(f_n)_i = \{(f_n)_i[1]\} \cup \{(f_n)_i
[2]\} \cup \dots \{(f_n)_i[j]\}$, where $i$ denotes the packet index in
flow $f_n$, $j$ denotes the byte-offset in packet $i$, and $\{(f_n)_i[j]\} \in
[0,255]$. Next, we join these packet sequences, delimiting them with a custom 
\texttt{<|pkt|>} special token to form a training sample. Finally, we prepend
each sample with a \texttt{<|LABEL|>} special token that
denotes its traffic type.
Thus, $\texttt{sample} = 
\{\texttt{<|LABEL|>}, \text{Seq}(f_n)_1, \texttt{<|pkt|>}, \text{Seq}(f_n)_2,
\ldots, \text{Seq}(f_n)_m \}$, where $m$ denotes the number
of packets in a given flow. We train a new tokenizer of the same base-type as
Mamba on these representations (GPT-NeoX-20B~\cite{black2022gpt}), adding
the \texttt{<|pkt|>} and \texttt{<|LABEL|>} special tokens.


\paragraph{Model implementation.}\label{ssub:model_training}
We use the open sourced model implementation released by Gu \etal~accompanying
the original Mamba work~\cite{gu2023mamba}. Specifically, this combines the
Mamba SSM described in the previous section with a gated multilayer perceptron
\cite{liu2021pay}, to form the Mamba block. Appendix~\ref{app:mamba_block}
provides further details on the architecture and its functionality. The Mamba
block operates in a causal (left-to-right) manner, and trains using the common
next token prediction task evaluated via cross entropy loss.

\paragraph{Trace generation.} \label{ssub:generation}
Our model generates synthetic network traces by taking in two arguments:
a generation \textit{seed} and \textit{length}.
The seed is a sequence consisting of a flow's label token, and sequence of raw
bytes that comprise its first packet (\eg, \texttt{<|twitch|> 205 68...}), and
is equivalent to a start token/prompt in NLP generative models. The generated
length dictates the maximum number of tokens output by the model. The output of
the model resembles the format of the training data, a sequence of raw
bytes of packets in decimal format, with each packet delimited by a 
\texttt{<|pkt|>} special token. We then convert the raw, string-based output
to binary format for use and evaluation as a PCAP file.
\section{Preliminary Evaluation}\label{sec:eval}

We assess the effectiveness of our approach by examining the 
\textit{packet-level, fine-grained generation quality} of traces generated by
our model, evaluated through:
\begin{enumerate}
	\item Flexibility to generate traces of varying lengths without quality degradation,
	\item Statistical resemblance of synthetic traces to real network data, and
	\item Empirical verification of the model's learning capability rather than memorization.
\end{enumerate}
We describe the datasets used for this preliminary
evaluation, implementation specifics, training recipes, and results
below.

\subsection{Experiment Setup}
\paragraph{Dataset Description.}
For generation quality assessment, it is essential to use network trace datasets
thoroughly cleaned to reduce noise, \ie, unintended traffic capture.
Statistical comparisons between real and synthetic traffic are more
credible and less variable with clean datasets,
which also enhance the effectiveness of traffic analysis tools in downstream
applications.
However, many public datasets are unsuitable due to high noise levels.
For our experiment, we use three representative, labeled datasets
previously used in similar studies for synthetic trace generation, comprised
of video streaming~\cite{bronzino2019inferring}, video conferencing
\cite{macmillan2021measuring}, and social media data
\cite{jiang2023ac}, as detailed in Table~\ref{tab:dataset_overview}.
Each dataset contains traffic from its respective application domain,
divided into flows in PCAP format.
We further filter these flows to ensure their
relevance to the labeled application domain
by extracting relevant DNS queries, resolving them to IP addresses,
and filtering the packets based on these addresses.
We aggregate these datasets to form a \textit{service-classification dataset},
which we use as the training dataset for our model.

\begin{table}[t!]
    \centering
	\resizebox{\columnwidth}{!}{
    \begin{tabular}{lcrr}
    \toprule
    \multicolumn{2}{c}{\textbf{Content Type}} & \multicolumn{2}{c}{
    \textbf{Size}} \\
    \cmidrule(lr){1-2}
    \cmidrule(lr){3-4}
    \textbf{Macro-Label} & \textbf{\# Micro-Labels} & \textbf{Raw} & 
    \textbf{Flows} \\
    \midrule
    Video Streaming~\cite{bronzino2019inferring} & 4 & 6.36 GiB & $10{,}032$\\
    Video Conferencing~\cite{macmillan2021measuring} & 3 & 17.36 GiB & $13{,}911$ \\
    Social Media~\cite{jiang2023ac} & 3 & 5.40 GiB & $3{,}896$\\
    \bottomrule
    \end{tabular}}
        \caption{Service-classification dataset overview.}
        \label{tab:dataset_overview}
\end{table}
 	
\paragraph{Pre-training Configuration.}
We pre-train our model on this service-classification dataset using a single
NVIDIA A40 48GB PCIe GPU, with gradient clip value of $1.0$ and the AdamW
optimizer with learning rate $5e-4$. We leave all other optimizer values as
their defaults ($\beta = (0.9, 0.999)$, $\epsilon = 1e-8$, weight decay $=
1e-2$). We use the same configuration for dimension ($768$) and number of layers
($24$) as the smallest $125$ million parameter pre-trained Mamba. We extend the
base Mamba tokenizer to include the delimiting (\texttt{<|pkt|>}) and ten label
special tokens for the traffic in the dataset. We train the generation model
following the process discussed in Section~\ref{ssub:preprocessing}, first
splitting all PCAPs into their comprising flows, parsing them into decimal
representations, and finally tokenizing these inputs. We run pre-training using
batch size of one, with maximum sequence length of $50{,}000$ tokens (the
largest possible given our hardware and configuration; training samples longer
than this length are truncated), for $50$ epochs, until cross-entropy loss
converges at $1.32$ nats. Here, we minimize batch size and maximize per sample
maximum sequence length, to have the model learn from as large of individual
contexts of sequential packets in a flow as possible.

\subsection{Evaluating Generation Quality}
Our evaluation focuses on our model's ability to accurately reproduce
intra-packet and intra-flow dependencies observed in real traces.
Specifically, we aim to generate synthetic traffic that closely resembles
real traffic dynamics at the packet level.
The
conventional empirical metric for assessing this quality, as noted in prior
studies \cite{jiang2024netdiffusion, yin2022practical, qu2024trafficgpt},
involves measuring the statistical similarity between the raw values of
generated packet header fields and those from real traces.
Below, we detail and discuss our generation process, benchmarks, and comparative results.

\paragraph{Conditioned Trace Generation.}
We generate synthetic network flows by using the first packet from each flow in
the service-classification dataset as the seed to create the corresponding
synthetic traces, as depicted in Figure~\ref{fig:pipeline}.
Recall that each packet begins with its corresponding label special token (\eg,
\texttt{<|twitch|>}, \texttt{<|youtube|>}, \texttt{<|zoom|>}) indicating the
trace type, to indicate to the model the the type of traffic to generate.
Therefore, a typical seed may appear such as \texttt{<|twitch|> 160 206 ...
<|pkt|>}, providing contextual cues for accurate synthetic trace generation.
We set the generation length at 10 tokens beyond the real tokenized
trace length to ensure the synthetic versions have similar packet counts
while allowing flexibility for variations and adjustments, such as truncating
malformed packets.

\paragraph{Benchmark Setup.}
We first benchmark output traces generated by our model against representative SOTA
network trace generators:
NetShare~\cite{yin2022practical} for coarse-grained and
NetDiffusion~\cite{jiang2024netdiffusion} for fine-grained.
Both models are trained on the service-classification dataset,
producing corresponding synthetic traces.
For NetDiffusion, a post-generation correction
heuristic is required to ensure semantic coherence in generated traces,
which our approach \textit{does not} need. We report the NetDiffusion results
after applying the the correction heuristic.
We also use the statistical similarity metrics reported by TrafficGPT
to compare against the latest transformer-based architecture for fine-grained trace generation.
Our comparison is based on their published results, as TrafficGPT is not open source
and uses different data. 
Note that TrafficGPT focuses on a
limited set of features (\eg, IP addresses, ports), whereas the benchmark for
NetShare, NetDiffusion, and our model consider similarity across \textit{all}
header values of the generated flow.
Thus, although we cannot retrain TrafficGPT on our dataset, we acknowledge that
its reported results likely showcase its optimal performance.

\begin{table}[tb]
  \centering
  \resizebox{\columnwidth}{!}{
    \begin{threeparttable}
      \begin{tabular}{l r r r r}
        \toprule
        \textbf{Data Format} & \textbf{Generation Method} & \textbf{JSD} & \textbf{TVD} & \textbf{HD} \\
        \midrule
        \multirow{2}{*}{NetFlow} & Random Generation & 0.67 & 0.80 & 0.76 \\
                                 & NetShare & 0.16 & 0.16 & 0.18 \\
        \midrule
        \multirow{4}{*}{Raw PCAP} & Random Generation & 0.82 & 0.99 & 0.95 \\
                                   & NetDiffusion$^\dagger$ & 0.04 & 0.04 & 0.05 \\
                                   & TrafficGPT\tnote{*} & 0.16\tnote{*} & N/A & N/A \\
                                   & \textbf{Ours} & \textbf{0.02} & \textbf{0.01} & \textbf{0.02} \\
        \bottomrule
      \end{tabular}
      \begin{tablenotes}
        \footnotesize
        \item[$\dagger$] Post-generation correction applied; *
        Results as reported in \cite{qu2024trafficgpt}.
      \end{tablenotes}
    \end{threeparttable}
  }
  \caption{Statistical similarity versus real traces (packet-level,
  fine-grained generation quality evaluation).}
  \label{tab:statistical_similarity}
\end{table}

\paragraph{Generated Trace Length.}
We verify that our model can generate long traces while
maintaining high quality. We observe the longest trace generated by our model
has length of $1{,}081$ packets ($65{,}943$ tokens), improving slightly
on NetDiffusion's max trace generation length ($1{,}024$), and
is $5.5$ times longer than TrafficGPT's max sequence generation length($12
{,}032$ tokens). Table~\ref{tab:packet_stats} in the Appendix provides further
overview of the length of traces generated by our model.

\paragraph{Statistical Similarity.}
We assess the statistical similarity between the generated traces and their
respective ground truth representations using the average Jensen-Shannon
Divergence (JSD), Total Variation Distance (TVD), and Hellinger Distance (HD)
metrics across all header field values. Lower scores in these metrics represent
higher statistical generation fidelity compared to the real traces. To ensure
uniformity across comparisons, all synthetic and real traces are converted to a
one-hot encoded binary representation using the nPrint format~\cite{nprint}.
We include a crude upper-bound benchmark as a baseline by
synthesizing all values randomly, without any model training or heuristic
application.

Table~\ref{tab:statistical_similarity} shows that
our model distinguishes itself by achieving the lowest JSD, TVD, and HD
scores among all evaluated network trace generators.
Compared to the coarse-grained NetShare generator,
we generate finer-grained raw traffic with closer
statistical resemblance to real network traffic,
achieving consistently lower metric scores,
even though NetShare generates simpler NetFlow attributes
that are more likely to achieve higher statistical similarity.
Further, when
comparing our model to the fine-grained NetDiffusion method, our approach still
leads by a notable margin in all three statistical similarity metrics.
This is particularly significant considering these results are
based on output without any post-generation correction, which NetDiffusion
requires to ensure semantic coherence. Considering reported results from
TrafficGPT, which also treats the trace generation task as a sequence modeling
problem, our approach still yields better fine-grained generation quality,
despite the fact that TrafficGPT's results are calculated on fields with
relatively fixed and consistent values. Overall, our model's performance
suggests a high degree of statistical fidelity to the original traffic patterns,
mimicing real-world behavior.

\paragraph{Learning versus memorization.}
To verify our model's learning capabilities, we perform one-to-one byte-wise
comparison of the first $100$ packets in each of our generated traces, and their
corresponding real-world packets, from which the generation seed is extracted
from.
Remarkably, on average, only one packet is identical, with the other $99$ being
different. The single identical packet can be explained by our trace generation
process, where we prompt our model with a seed consisting of the service label,
and first packet of a real-world flow. Common to other sequence generation
models, our model copies the seed as part of the output sequence. All other
generated packets have variances that separate them from their real-world
counterparts. Further, in these packets, the average percentage of differing
bytes is $6.57\%$. We analyze the distribution of these bytes (with detailed
results in Appendix Table~\ref{tab:pkt_byte_distr}) and find that they manifest
in fields non-essential to flow state, indicating our model can discern between
fields in the seed that should remain fixed (\eg, source, destination IP
addresses), and fields that when changed, will not disrupt communication.
These results show that our model effectively captures and generalizes
underlying network patterns, rather than memorizing specific data.
\section{Open Challenges}\label{sec:future}

\paragraph{Targeted generation and improving output flexibility.}
Training a Mamba SSM-based generation model using all
contents from all flow packets allows the model to learn the important
semantic relationships and underlying dependencies in general flows, evident
from the high levels of statistical similarity to real
traffic and observed packet byte distribution in our evaluation.
One useful future direction is to fine-tune our general model to create
models that can accurately generate the distinct phases or
components observed in authentic network traces.
As different stages of a network flow or session exhibit unique
patterns/behaviors, more targeted models could better capture the specific
intra-packet and intra-flow dynamics of these segments, improving
both the quality and flexibility of generated data.
This flexibility could be particularly useful for reducing generation overhead
for applications that need only a subset of flow data.

\paragraph{Temporal generation in synthetic traces.}
Our current model produces raw byte values across different network layers, but
does not generate temporal information, \ie, timestamps. Timestamps, determined
by packet capturing tools at the OS level, are not embedded in the raw packet
bytes and thus are not considered in the training/generation process.
Temporal information provides critical details for various tasks (\eg,
classification, attack detection).
Directly adopting the vanilla Mamba architecture for this
task (as done in this study) is insufficient; it is designed for
generating sequential discrete values, whereas temporal information is
continuous. Thus, to improve the practicality,
the Mamba/SSM architecture must be modified to generate
in-tandem both discrete packet-level values, and continuous
timestamps.

\paragraph{Better evaluation of synthetic trace quality and
utility.}
While statistical similarity is an important quality indicator,
it does not guarantee that generated synthetic traces are
practical and/or adhere to protocol constraints. To illustrate,
consider the IP version header, which in practice should only take on one of
two values: \texttt{0100} (IPv4) or \texttt{0110} (IPv6). A synthetically
generated value of \texttt{0101}, though statistically
close to valid values, would likely result in a dropped packet if
transmitted in the real-world.
Thus, it is crucial to further inspect these synthetic traces
to ensure their coherence. Finally, it is equally important to evaluate the
usefulness of these synthetic traces in common downstream tasks (\eg, training
and augmenting machine learning models on network traffic) and
towards facilitating existing heuristic-based analysis tools (\eg, anomaly
detection mechanisms).
\section{Conclusion}\label{sec:conclusion} In this work, we presented a new
approach to synthetic network trace generation, posing the task as a
unsupervised sequence generation problem using the Mamba state-space model
architecture. Our preliminary evaluation demonstrates that this method produces
synthetic traces with significantly higher statistical similarity to real-world
traffic compared to current state-of-the-art techniques. We outline future
improvements and applications for our model towards improving the realism and
utility of synthetic network traces, for use in different downstream tasks.
	\label{endOfBody}
	\bibliographystyle{plain}
	\bibliography{sigcomm_2024}
  \clearpage
  \appendix

\setcounter{figure}{0}
\setcounter{table}{0}
\setcounter{section}{0}

\section{Configuration}
The following section provides details referenced in the paper regarding
model architecture/training.

\subsection{Mamba Block}\label{app:mamba_block}

\begin{figure}[h]
  \centering
  \includegraphics[width=0.25\textwidth]{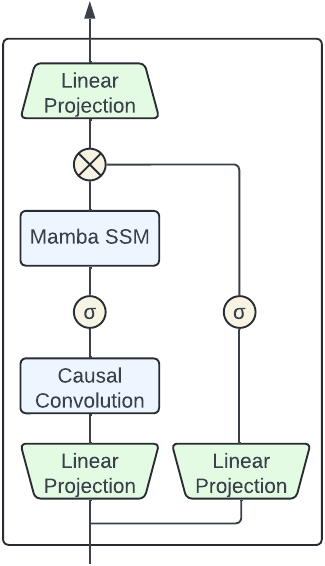}
  \caption{Architecture for the Mamba block, described in Section~
  \ref{sec:architecture}. $\sigma$ denotes the SiLU/Swish non-linear
  activation. $\otimes$ denotes element-wise multiply.}
  \label{fig:mamba_block}
\end{figure}

Figure~\ref{fig:mamba_block} depicts the Mamba block. Here, an input sequence
is linearly projected twice to the input dimension. One copy is passed through
a FFT-based causal convolution and SiLu non-linear activation
\cite{ramachandran2017searching}, before being used as input to the Mamba SSM.
This allows for efficient training in parallel via convolution, and
improves approximation of the true input distribution. The second copy passes
only through a SiLu activation, and ``gates'' the output of the Mamba
SSM from the first copy, \ie, using element-wise multiplication, enhances or
filters the Mamba SSM output. The gated output is then linearly projected back
to the original input dimension.

\section{Evaluation}
The following section provides additional details referenced in the paper
regarding model evaluation.

\subsection{Generation Length}


\begin{table}[h!]
  \centering
  \begin{tabular}{@{}lrrrrr@{}}
    \toprule
    \textbf{Metric} & \textbf{Max} & \textbf{Min} & \textbf{Mean} & \textbf{Std Dev} & \textbf{Var} \\ 
    \midrule
    Packets & 1081 & 4 & 123.17 & 230.57 & 53161.55 \\
    \bottomrule
  \end{tabular}
  \caption{Packet count statistics of generated flows.}
  \label{tab:packet_stats}
\end{table}

Table~\ref{tab:packet_stats} provides statistics for packet
length (in bytes) of packets in our generated synthetic traces.

\subsection{Comparing synthetic and real-world packet data}


\begin{table}[h!]
  \centering
  \begin{threeparttable}
    \begin{tabular}{lc}
      \toprule
      \textbf{Field} & \textbf{Average Change} \\
      \midrule
      TCP\_ack & 1982981477.85 \\
      TCP\_seq & 1475508827.26 \\
      TCP\_chksum & 24859.46 \\
      TCP\_sport & 23406.24 \\
      TCP\_dport & 23406.24 \\
      IP\_chksum & 6369.34 \\
      IP\_id & 2924.89 \\
      TCP\_window & 1136.37 \\
      IP\_len & 507.86 \\
      IP\_ttl & 11.52 \\
      TCP\_dataofs & 2.16 \\
      Raw\_load & 1.00 \\
      TCP\_options & 0.99 \\
      TCP\_flags & 0.86 \\
      Ether\_dst & 0.48 \\
      Ether\_src & 0.48 \\
      IP\_src & 0.48 \\
      IP\_dst & 0.48 \\
      Ether\_type & 0.00 \\
      IP\_version & 0.00 \\
      IP\_ihl & 0.00 \\
      IP\_tos & 0.00 \\
      IP\_flags & 0.00 \\
      IP\_frag & 0.00 \\
      IP\_proto & 0.00 \\
      IP\_options & 0.00 \\
      TCP\_reserved & 0.00 \\
      TCP\_urgptr & 0.00 \\
      \bottomrule
    \end{tabular}
  \end{threeparttable}
  \caption{Average change of representative header fields}
  \label{tab:pkt_byte_distr}
\end{table}

Table~\ref{tab:pkt_byte_distr} shows the average change in packet bytes of
synthetic traces and real-world data from our service-classification dataset.
Average change is calculated by comparing the fields of real and synthetic
packets, computing the absolute differences for each field, and averaging those
differences across all compared packets.
	
	
\end{sloppypar}
\end{document}